\documentclass[a4paper]{jpconf}
\usepackage{graphicx}
\usepackage{subcaption}

\PassOptionsToPackage{hyphens}{url}\usepackage{hyperref}

\begin{document}

\begin{center}
{Preprint accepted at CISBAT 2023 - The Built Environment in Transition, Hybrid International Conference, EPFL, Lausanne, Switzerland, 13-15 September 2023}
\end{center}

\title{Utilizing wearable technology to characterize and facilitate occupant collaborations in flexible workspaces}

\author{Kristi Maisha$^{1,2}$, Mario Frei$^1$, Matias Quintana$^{1,3}$, Yun Xuan Chua$^1$, Rishee Jain$^2$, Clayton Miller$^{1,*}$}
\address{$^1$ College of Design and Engineering, National University of Singapore (NUS), Singapore}
\address{$^2$ Urban Informatics Lab, Stanford University, USA}
\address{$^3$ Future Cities Laboratory Global, Singapore-ETH Centre, Singapore}

\ead{$^{*}$clayton@nus.edu.sg}

\begin{abstract}
Hybrid working strategies have become, and will continue to be, the norm for many offices. 
This raises two considerations: newly unoccupied spaces needlessly consume energy, and the occupied spaces need to be effectively used to facilitate meaningful interactions and create a positive, sustainable work culture. 
This work aims to determine when spontaneous, collaborative interactions occur within the building and the environmental factors that facilitate such interactions.
This study uses smartwatch-based micro-surveys using the Cozie platform to identify the occurrence of and spatially place interactions while categorizing them as a collaboration or distraction. 
This method uniquely circumvents pitfalls associated with surveying and qualitative data collection: occupant behaviors are identified in real-time in a non-intrusive manner, and survey data is corroborated with quantitative sensor data. 
A proof-of-concept study was deployed with nine hybrid-working participants providing 100 micro-survey cluster responses over approximately two weeks.
The results show the spontaneous interactions occurring in hybrid mode are split evenly among the categories of \emph{collaboration}, \emph{wanted socialization}, and \emph{distraction} and primarily occur with coworkers at one's desk.
From these data, we can establish various correlations between the occurrence of positive spontaneous interactions and different factors, such as the time of day and the locations in the building. 
This framework and first deployment provide the foundation for future large-scale data collection experiments and human interaction modeling. 

\end{abstract}

\section{Introduction}\label{ch:introduction}
Human-to-human interactions play an essential role in defining our society. 
Within the office, collaboration and interactions among employees are key determining factors for productivity and success \cite{Sonta2020-pd}. 
Specifically, unplanned, spontaneous interactions can provide opportunities for unexpected collaboration and discovery. 
Analyses of research laboratories in the US found that during tasks that involved creative problem-solving, discussions with coworkers were essential, and someone's successful performance could be correlated to the breadth of and easy access to their network \cite{Parkin2011-xp}. 
These benefits, however, are accompanied by the possibility of interaction being a distraction to one's workflow. 
The assessment of interactions is often accompanied by an analysis of spatial planning: an open plan may allow for more collaboration but also introduce noise and a lack of privacy, or the opposite for closed cubicles.

The COVID-19 pandemic has altered the paradigm of human interactions within workspaces. 
Price Waterhouse Cooper's 2021 US Remote Work survey found that only 21\% of employers felt that employees needed to be in the office full time, and 55\% of employees wanted to work remotely at least three days per week, but 87\% of employees still felt that the office space was essential for collaboration \cite{pricewaterhousecoopers_business_nodate}. 
Space utilization and teamwork among employees are significant concerns for companies.
A recent study shows that when comparing interactions conducted remotely rather than in person, a switch to remote work often leads to a smaller collaboration network, indicating less knowledge transfer \cite{Yang2022-gj}. 
In-person and synchronous communication (such as video calls) is better for more complex information, reaching a common understanding, and community building, but when remote, most people switch to asynchronous communication (messages and emails). 
This study also indicated that a shift to remote work could lead to reduced productivity and innovation, leading to the following questions:
Given the hybrid work dynamics in a post-COVID-19 world, how can we identify quality interactions among employees? 
How can existing building stock be accurately optimized to support quality occupant interactions?

\subsection{Identifying interactions in the workplace}
The occurrence of occupant interactions has previously been studied and predicted using sensor data. 
The activity state of the plug-in devices at one's desk can be used to infer whether someone is at their workstation \cite{Sonta2020-pd}. 
When more than one occupant is away from their desk, interactions can occur, creating a model that reflects potential employee interactions.
Another method uses environmental data, such as noise levels, CO\textsubscript{2} levels, or temperature, to predict the occurrence of interactions \cite{Ghahramani2018-ll}.
However, the occurrence of an interaction in these studies does not indicate whether the interaction was meaningful.
Optimizing our buildings to best support quality interactions and best advise people's habits and behaviors requires a more nuanced and subjective understanding of how each individual perceives the impact of their interactions. 
In other words, humans must become the sensor and provide data on their opinions and feelings. 
The smartwatch application Cozie (\href{https://cozie-apple.app/}{cozie-apple.app/}) was developed as a data collection platform that uses micro-ecological momentary assessments (micro-EMAs) to collect right-here-right-now user perceptions of their environment \cite{Jayathissa2019-kg}. 
In the context of occupant interactions, this means a near-immediate recall of recent interactions and their associated value. 
Cozie has also been effectively used to assess occupant experiences with thermal comfort, movement, infectious disease risk, privacy, noise distractions, and to nudge occupants to make decisions \cite{Miller2022-sj, Quintana2021-ka, Miller2022-dy}. 

This work presents and pilots a method to use the Cozie app and surveying to establish a pattern of interactions within office spaces. 
This method uniquely provides the ability to identify and differentiate interactions based on the value they provide to a person's work and personal life. 
This data allows for correlations between collaborative, spontaneous interactions and different factors, such as the time of day and the locations in the building. 
In addition, we understand which general topics are most prevalent (work, personal life, etc.) as well as the perceived value of these interactions from the occupants themselves.

\section{Methodology}\label{ch:Methodology}
\begin{figure}[!htb]
\begin{center}
\includegraphics[width=1\textwidth, trim= 0cm 0cm 0cm 0cm,clip]{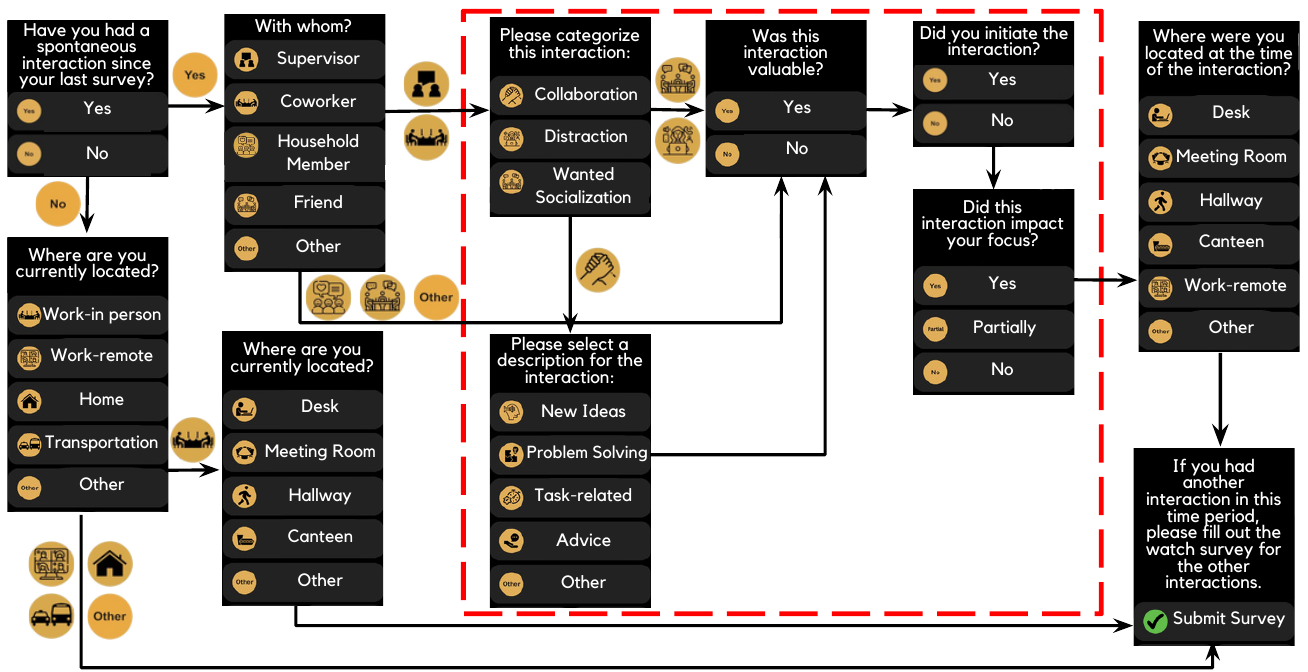}
\caption{Occupant interaction micro-survey flow overview. 
The arrows indicate the flow of questions. The questions within the red dashed box assess the user's subjective experience.}
\label{fig:qflow}
\vspace{-20pt} 
\end{center}
\end{figure}

Four data collection methods were developed for this study: an onboarding survey at the start of the study, micro-EMAs surveys, an end-of-day survey, and an exit interview that was tailored to each participant's individual survey responses. The onboarding survey aims to gain a baseline understanding of each participant's schedule and work patterns as well as their feelings regarding the value of in-person work. 
Figure \ref{fig:qflow} shows the micro-EMA survey developed to assess whether a spontaneous interaction occurs and the circumstances around that interaction. 
The questions within the red dashed box gather data on the subjective details of the interaction, while the other questions identify the objective characteristics.  
These subjective questions aim to determine whether the interaction positively impacts the participant's work performance and are being compared to identify the most effective language. 
The final screen of the micro-survey prompts the user to complete the flow again for any other spontaneous interactions that occurred after the previous submission. 
The set of responses given at one time is referred to as a micro-survey cluster. 
The end-of-day survey, designed to take less than five minutes, validates the interaction count found through the micro-EMA surveys and asks the participant to reflect on the interactions' overall value and general topics. 
The three question sets described above were reviewed by external researchers in the field. 
Finally, in the exit interview, participants are asked to provide their thought processes as they complete the surveys and explain their feelings on the accuracy of the questions. 
This step included discussions of the participant's definition of subjective terms, such as \emph{valuable} and \emph{quality}, and discussions of their confidence in interaction recall when answering the micro-EMA and end-of-day survey.

\subsection{Experimental deployment}
This pilot study aims to assess the efficacy of the developed question flow and surveys and determine the relevance of the collected data.
Nine participants were selected to wear an Apple smartwatch for approximately two weeks and submit at least 100 micro-survey clusters. 
All participants had flexible, hybrid work schedules with a workspace in the same Singaporean office building, though potentially in different spaces. 
Reminders to complete a micro-survey cluster were given every hour on their iPhone and Apple Watch from 9:00 AM to 7:00 PM. 
At 7:00 PM, participants were also reminded to complete the end-of-day survey. 
Beyond the participant responses, the experiment collected data on heart rate, step count, and noise levels from the Apple Watch's built-in sensors.

\begin{figure}[!b]
\begin{center}
\includegraphics[width=1\textwidth, trim= 0cm 0cm 0cm 0cm,clip]{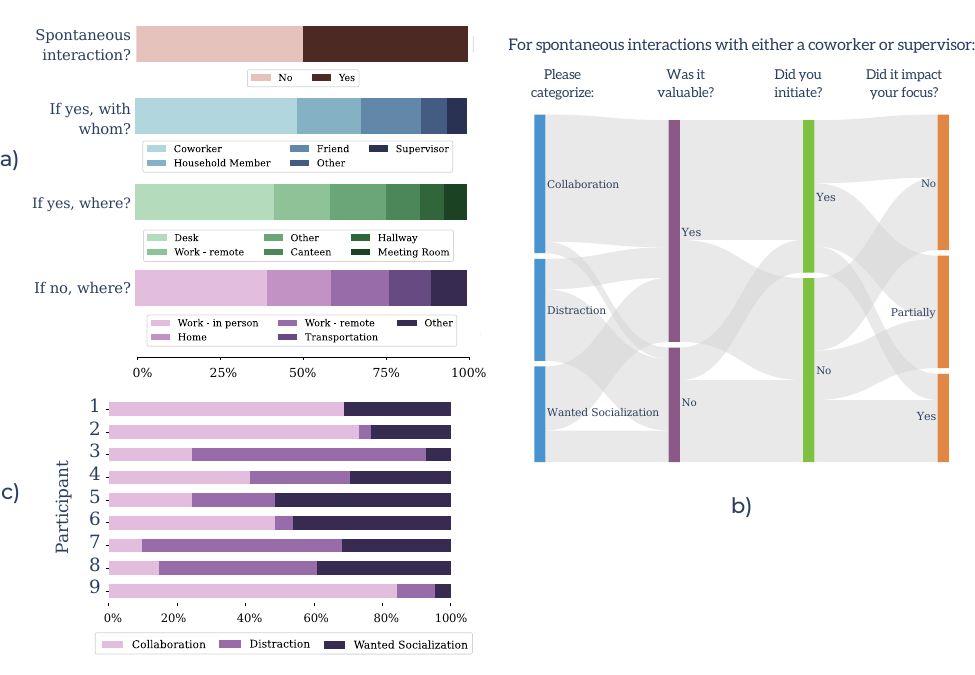}
\caption{Results from the micro-survey data a)  the distribution of objective characteristics for the interactions.  b) Sankey plot displaying the flow of question responses to the subjective questions and c) the individual responses to categorizing the interactions exemplifies the different work and response patterns that participants may have }
\label{fig:results}
\end{center}
\end{figure}

\section{Results and discussion}\label{ch:Results and Discussion}
The micro-survey was generally effective in contextualizing interactions, particularly for the objective questions. 
Figure \ref{fig:results}a displays the response breakdown for the objective questions. 
About half of the micro-survey responses were given when a spontaneous interaction had occurred.
Out of these, the primary response of \emph{with whom} and \emph{where} the interactions occurred was with coworkers at the participant's desk.
This indicates that a significant pool of the responses is relevant to an occupant's work experience in the office.
Participants felt confident that the recall period of one hour was appropriate for remembering interactions easily. 
Additionally, they felt that most aspects of the interaction and its impact on them were covered with the survey questions. 

Figure \ref{fig:results}b shows the results of the subjective micro-survey questions (those outlined with a red dashed line in Figure \ref{fig:qflow}) that were designed to address how the interaction affected the participant. 
These results demonstrate that a significant portion of collaborations are deemed valuable, and a significant portion of distractions are not, as expected, but the wanted socialization category is more evenly split.
Of the 90 \emph{Wanted socialization} interactions, 51\% were said not to impact the participant's focus, while the remaining 49\% partially or fully impacted the participant's focus.
These near-even splits indicate that the question of categorizing the interaction is not redundant and provides a unique dimension of the interaction's impact.
The results also show that whether the participant initiated the interaction or not has very little correlation to whether or not the interaction was valuable or if the interaction impacted their focus.
It was expected that if an interaction was deemed valuable, the participant would not consider it to have an impact on their focus.
Out of the 211 valuable interactions, half of them (41\%) had no impact on the participants' focus, while the other half (59\%) interactions at least partially impacted their focus, meaning that there is actually not a strong correlation between value and impact on focus. 

For the end-of-day survey (results not visualized), participants felt varying levels of confidence in answering the questions.
For some, recalling interaction counts for the entire day was difficult. 
The split of topics was usually divided somewhat evenly between work and personal life, with a slightly higher percentage of work topics. 
One participant indicated that they defaulted to 60-70\% of interactions being work-related and the remaining being personal life-related unless something significantly different occurred during the day.
The participants did, however, indicate that the end-of-day survey reflected the overall interaction patterns of the day, which the participants felt was personally interesting.
A few of the participants also indicated that their perception of the interactions when completing the  end-of-day survey differed from their impression at the moment when completing the micro-survey.
Data and code from this deployment are found in the following open GitHub repository: \url{https://github.com/buds-lab/occupant-interactions-workspaces}

\subsection{Suggestions for methodological improvement for future studies}
In the onboarding session, providing a more in-depth understanding of what each question is asking for could be valuable. 
While the goal is to obtain information on the experience of the participant, participants were not always confident about their selections, which could lead to more frustration during the experimental process. 
In the exit interview, participants were asked to consider an interaction that they had experienced and describe their thought process at each question. 
Including this exercise in the onboarding would be valuable, where the participant can demonstrate how they plan to approach each question. 
This would allow the researcher to determine any glaring interpretation issues as well as give the participant confidence in their understanding, making the survey process smoother and less cognitively intensive. 
A similar walkthrough with the end-of-day survey could also eliminate some of the confusion and provide the participants with more substantial reasoning behind each question. 
 
For the questions themselves, multiple participants indicated an interest in a question asking about the duration of the interaction. 
This could be a better indicator of the impact on focus or value depending on the interaction categorization while also providing a benchmark for what extent of conversation is considered an interaction. 
Additionally, a phrasing change should be considered for the option of \emph{Home} when asking about the location to reflect better the intended meaning of leisure time in one's home and \emph{Household member} to separate out further one's partner, which is currently usually categorized under other by the participants. 
Beyond the pain point of terminology, some participants expressed that the working hours required by the study were a bit bothersome as they started work later or ended earlier. 
Changing the parameters of each participant's work hours could also reduce potential frustrations. 
Another suggestion is related to the fact that terms such as \emph{valuable}, \emph{focus}, and \emph{quality} are defined differently from person to person. 
While it is indicated above that some level of clarification and defining would be beneficial for the participant, it is also necessary to balance the need for an authentic appraisal of the interaction's impact, which could be lost with over-prescription. 

It is also important to note that the definitions of such terms vary depending on the participant's mood or current schedule. 
For instance, to the question, \emph{Did this interaction impact your focus?}, one participant never answered \emph{yes} despite marking the interaction as a distraction because the participant felt that they were currently not in a state of intense focus. 
Other participants indicated that whether an interaction was \emph{valuable} could vary depending on the mood that they were in that day. 
In a similar vein, when categorizing the interactions, some participants considered almost all non-work-related interactions to be distractions.
In contrast, other participants felt that virtually no interaction was purely a distraction and rarely selected it. 
Figure \ref{fig:results}c demonstrates these participant differences. 
Participants \#1, 2, and 6 rarely, if ever, selected distraction, as they stated that some conversation with colleagues is beneficial, and they had a higher threshold for what might be distracting.
In contrast, Participants \#5, 7, and 8 have relatively fewer spontaneous collaborations and tend to categorize non-work-related interactions as distractions more readily.
These different profiles lend credence to the inclusion of multiple answer choices (i.e., distraction and wanted socialization) and multiple layers of subjective questions.

\section{Conclusions}
This study details developing and testing a method for identifying and assessing interactions during the workday, particularly within the physical office space. 
A question flow was established and implemented into the Cozie app to collect information on where and with whom an interaction is occurring, as well as the impact of the interaction on the occupant's personal and work life. 
An onboarding survey, end-of-day surveys, and exit interviews supported this question flow. 
The method was tested with nine participants working within the same office building and was shown to capture the quantity and impact of interactions successfully. 
It also highlighted the differences in how people interpret that impact, which should be incorporated into the experimental design of more significant future deployments. 
In future work, the method can be aided with other sensor-based data, such as noise levels or workspace occupancy, to assess better when interactions are occurring.

\section{Acknowledgements}\label{ch:Acknowledge}
This research has been supported by the following Singapore Singapore Ministry of Education (MOE) Tier 1 Grants: A-0008305-01-00, A-0008301-01-00, and A-8000139-01-00. Kristi Maisha gratefully acknowledges financial support for this research by the Fulbright U.S. Student Program, which is sponsored by the U.S. Department of State. 

\section*{References}
\bibliographystyle{elsarticle-num} 
\bibliography{references} 

\begin{thebibliography}{1}
\expandafter\ifx\csname url\endcsname\relax
  \def\url#1{\texttt{#1}}\fi
\expandafter\ifx\csname urlprefix\endcsname\relax\def\urlprefix{URL }\fi
\expandafter\ifx\csname href\endcsname\relax
  \def\href#1#2{#2} \def\path#1{#1}\fi

\bibitem{Sonta2020-pd}
A.~Sonta, R.~K. Jain, Learning socio-organizational network structure in
  buildings with ambient sensing data, Data-Centric Engineering 1 (2020) e9.

\bibitem{Parkin2011-xp}
J.~K. Parkin, S.~A. Austin, J.~A. Pinder, T.~S. Baguley, S.~N. Allenby,
  Balancing collaboration and privacy in academic workspaces, Facilities
  29~(1/2) (2011) 31--49.

\bibitem{pricewaterhousecoopers_business_nodate}
{PricewaterhouseCoopers},
  \href{https://www.pwc.com/us/en/services/consulting/business-transformation/library/covid-19-us-remote-work-survey.html}{Business
  needs a tighter strategy for remote work}.
\newline\urlprefix\url{https://www.pwc.com/us/en/services/consulting/business-transformation/library/covid-19-us-remote-work-survey.html}

\bibitem{Yang2022-gj}
L.~Yang, D.~Holtz, S.~Jaffe, S.~Suri, S.~Sinha, J.~Weston, C.~Joyce, N.~Shah,
  K.~Sherman, B.~Hecht, J.~Teevan, The effects of remote work on collaboration
  among information workers, Nat Hum Behav 6~(1) (2022) 43--54.

\bibitem{Ghahramani2018-ll}
A.~Ghahramani, J.~Pantelic, C.~Lindberg, M.~Mehl, K.~Srinivasan, B.~Gilligan,
  E.~Arens, Learning occupants' workplace interactions from wearable and
  stationary ambient sensing systems, Appl. Energy 230 (2018) 42--51.

\bibitem{Jayathissa2019-kg}
P.~Jayathissa, M.~Quintana, T.~Sood, N.~Nazarian, C.~Miller, Is your clock-face
  cozie? a smartwatch methodology for the in-situ collection of occupant
  comfort data, J. Phys. Conf. Ser. 1343~(1) (2019) 012145.

\bibitem{Miller2022-sj}
C.~Miller, R.~Christensen, J.~K. Leong, M.~Abdelrahman, F.~Tartarini,
  M.~Quintana, A.~M. M{\"u}ller, M.~Frei, Smartwatch-based ecological momentary
  assessments for occupant wellness and privacy in buildings, in: Indoor Air
  2022 - 17th International Conference of the International Society of Indoor
  Air Quality \& Climate, 2022.

\bibitem{Quintana2021-ka}
M.~Quintana, M.~Abdelrahman, M.~Frei, F.~Tartarini, C.~Miller, Longitudinal
  personal thermal comfort preference data in the wild, in: Proceedings of the
  19th {ACM} Conference on Embedded Networked Sensor Systems, SenSys '21,
  Association for Computing Machinery, New York, NY, USA, 2021, pp. 556--559.

\bibitem{Miller2022-dy}
C.~Miller, Y.~X. Chua, M.~Frei, M.~Quintana, Towards smartwatch-driven
  just-in-time adaptive interventions (jitai) for building occupants, in:
  Proceedings of the 9th ACM International Conference on Systems for
  Energy-Efficient Buildings, Cities, and Transportation, BuildSys '22,
  Association for Computing Machinery, New York, NY, USA, 2022, p. 336–339.
\newblock \href {https://doi.org/10.1145/3563357.3566135}
  {\path{doi:10.1145/3563357.3566135}}.

\end{thebibliography}

\end{document}